# Uncovering the structure of the French media ecosystem

Jean-Philippe Cointet[1,2], Dominique Cardon[1], Andreï Mogoutov,[1] Benjamin Ooghe-Tabanou[1], Guillaume Plique[1] and Pedro Ramaciotti Morales[1]

## Abstract:

This study provides a large-scale mapping of the French media space using digital methods to estimate political polarization and to study information circuits. We collect data about the production and circulation of online news stories in France over the course of one year, adopting a multi-layer perspective on the media ecosystem. We source our data from websites, Twitter and Facebook. We also identify a certain number of important structural features. A stochastic block model of the hyperlinks structure shows the systematic rejection of counter-informational press in a separate cluster which hardly receives any attention from the mainstream media. Counter-informational sub-spaces are also peripheral on the consumption side. We measure their respective audiences on Twitter and Facebook and do not observe a large discrepancy between both social networks, with counter-information space, far right and far left media gathering limited audiences. Finally, we also measure the ideological distribution of news stories using Twitter data, which also suggests that the French media landscape is quite balanced. We therefore conclude that the French media ecosystem does not suffer from the same level of polarization as the US media ecosystem. The comparison with the American situation also allows us to consolidate a result from studies on disinformation: the polarization of the journalistic space and the circulation of fake news are phenomena that only become more widespread when dominant and influential actors in the political or journalistic space spread topics and dubious content originally circulating in the fringe of the information space.

## Overview:

With the emergence of social networks, the circulation of digital information has profoundly changed the overall structure of public media spaces. The flow of digital information has contributed to a series of mostly harmful phenomena such as filter bubbles (Pariser, 2011 ; Bruns, 2019), disinformation (Allen & al., 2020 ; Del Vicario et al., 2016), political polarisation (Bail, 2021 ; O'Hara and Stevens, 2015), hate speech (Marwick, 2021 ; Silva et al, 2016) and, more broadly, the weakening of the authority of traditional gatekeepers (Napoli, 2019 ; Wu, 2019). The consequences of the digital transformation of the public space on traditional media, the shaping of public opinion as well as on democracy have been extensively studied (Persilly, Tucker, 2020). However, changes in public spaces are not only the result of the emergence of a digital infrastructure for the production and exchange of information (Aral, 2020). They depend on the history, organizational structure and dynamics that are specific to each national context. The aim of this article is to show the historical, social and political specificity of the organization of the French media space, by comparing it to the phenomenon of media polarization observed in the United States. More precisely, we identify striking differences when comparing with similar methods the structure of the  French digital mediaspace with the structure observed by Benkler et al. (2018) in the case of  the United States. This comparison highlights the importance of social and political factors in the organization of national information systems, even if they are all subject to the global effects of the expansion of digital social networks.

---

[1] Sciences Po, médialab, 27 rue St Guillaume, 75337 Paris Cedex 07, France
[2] jeanphilippe.cointet@sciencespo.fr

In their research, Benkler et al. (2018) show that the polarization of political and media space has been so forceful that it has set a boundary within the central media space. Outlets such as Fox News and Breitbart, which have very large audiences, have become the relayers and translators of issues and disinformation produced in the peripheries of the American public space, especially on far-right websites. A (propaganda) feedback loop has developed in the United States, allowing information produced by extremist niches to gain visibility at the centre of the public space (Kaiser et al., 2020), while producing a very strongly polarized split between the major American media. Using digital methods that provide a large-scale overview of the digital public space, our study shows that in 2019, in the French case, we do not observe this disconnection within the space of professional media. This may explain why false information has less of an opportunity to enter the general political agenda.

The growing availability of online data from various API points increasingly allows us to model the digital public space as an ecosystem consisting of the interplay of various media platforms. Despite the technical and modeling challenge it represents, cross-platform research has become mandatory for capturing the realities of the contemporary media experience (Bode and Vraga, 2018; Boyd and Crawford, 2012). This ambition motivates the design of this research which considers three distinct layers of the digital public space to model the French media system distinguishing on top, the layer of media websites contributing to the production of news, and below, two layers composed of online conversation about these French news stories on Twitter and on Facebook (see Fig. 1). Each layer corresponds to a different modeling of the structure of the media space. In the top layer, the relationships between media are built from the hyperlinks between articles written by journalists. The representation of the media space is akin to a measure of *authority* produced by journalists' judgments toward their colleagues. The second layer of the model is built thanks to the shares of media articles by Twitter users. The corresponding media space representation is that of *influence*: share counts measure the influence of different news outlets within the specific social world of Twitter. Finally, at the third layer of this model, we observe how Facebook users share links to articles in their daily *conversations*.  We argue that Facebook because of its more conversational style and its less biased socio-demographic distribution than Twitter (Mislove et al. 2011) is likely to exhibit distinct patterns that are useful to explore the structure of online news consumption phenomena. This three-story mode offers a comparative vantage point with the work of Benkler et. al. (2018) on the US media ecosystem.

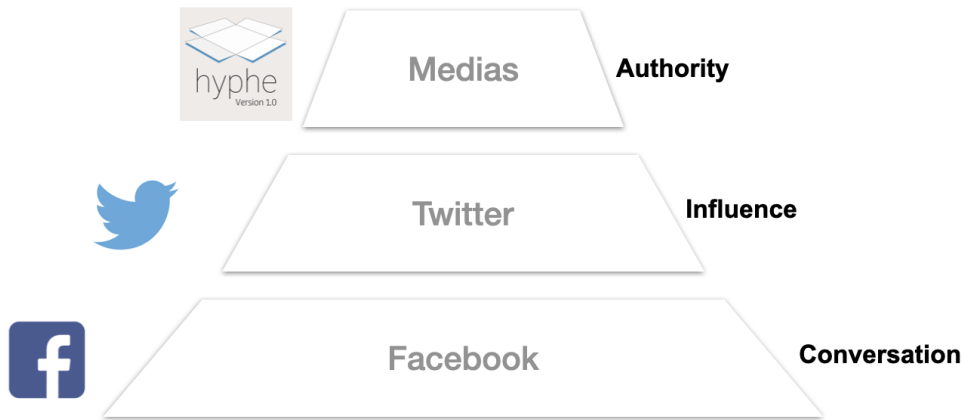

**Fig 1.** The three-level model of the digital media space - we distinguish between the media website producing news stories online, and Twitter and Facebook where news stories are being shared, commented on and discussed.

# Rationale and main findings

Our empirical study spans over eighteen months, from April 2018 to November 2019. Our corpus is defined by a closed set of 421 prominent media outlets which we identified both from pre-existing listings and from structural properties of the hyperlink network. We then used the advanced open source web crawling solution Hyphe (Ooghe-Tabanou et al., 2018) to extract hyperlinks connecting one media outlet with another. The organization of the resulting directed network is analyzed using a stochastic block model (Peixoto, 2014). Its block structure shows how mainstream media establishes a very clear frontier with counter-informational space thanks to a strict discipline in their hyperlink distribution strategy. We then turn to Twitter and Facebook to measure how many times each story coming from each block was shared. Statistics show that the attention is also very heterogeneously distributed on both social networks, with contentious sources hardly receiving any attention.

Moreover, the French political system cannot be reduced to a one-dimensional opposition between liberals and conservatives. As a matter of fact, the renewal of the traditional system of opposition between right and left wing parties is being increasingly questioned in the public debate ever since the election of Emmanuel Macron, who made his way to the Elysée with an ideologically blurred program (Strudel, 2017), and who was famously epitomized by his tendency to abuse of rhetorics refusing traditional left/right oppositions ("en même temps"). Beyond this recent evolution (and because both the French electoral model as well as its political system are not simply bipartisan) it was impossible for us to assign each media a political score in a straightforward way. As a consequence, we use latent ideology scoring methods to embed Twitter users in a complex two-dimensional space and later compute average political positions at the story and then the media level. In turn, this latent space allows us to measure the distribution of the political orientation of our full set of news outlets. In stark contrast with the afore-mentioned study, we do not observe the same level of polarization of far right sources.

The French media ecosystem seems to have self-disciplined in order to protect mainstream media from the intrusion of potentially harmful players in the ecosystem. On Twitter we do not observe any over-representation of extreme-right or extreme-left content. That being said, it does not mean that alternative media outlets at the periphery of the core set of outlets forming the "establishment journalism" do not exist. They are simply not really visible in routine political situations. On the other hand, when a long and important social movement such as the *Gilets Jaunes* occupied the news from October 2018 to June 2019, Facebook emerged as the main medium for the expression and representation of the mobilization. The prevalence of Facebook can be explained, in part, by the social homogeneity of the central media space and its remoteness from the grievances and demands from the *Gilets Jaunes* (Ramaciotti Morales et al., 2021).

# Data

In line with the multi-layered nature of the digital media space, our data collection strategy consists in 5 different datasets which are collected at 3 different layers:
- at the top level, we construct the hyperlinks network that links French media the one with the other (D1);
- from the Twitter layer, we collect every single tweet citing our predefined set of news sources over the same period (D2), and we also estimate the ideological position of nearly 350k Twitter users following each at least 3 members of the French Parliament (D3);
- finally, we estimate the popularity of news stories on Facebook thanks to the Condor dataset (D4).

## Media Websites

We first set the perimeter of French information media. We defined the list using a first set of sources identified thanks to authoritative sources such as Wikipedia or Le Monde's decodex[3]. This first list of sources was then used as seeds from which we "snow-balled" to identify other prominent sources. We used Hyphe (Ooghe-Tabanou et al., 2018) to perform this type of crawl. We stopped the process when the crawl hit depth 3. The resulting network was then pruned to only keep nodes with at least one incoming and out-going link. Overall, the procedure allowed us to capture 421 media sources. The final list, although seeded using a pre-existing list, is hardly dependent on this first decision, since influential websites are naturally "re-discovered" in the process because they tend to be linked by others. The dataset D1 is a network of news sources composed of 421 nodes and 20992 directed edges. We did not exploit the information about the number of links in-between media as the potential duplication of URLs referring to the same story complicated the interpretation of such information. The dataset is fully documented and publicly accessible online.[4]

---

[3] https://www.lemonde.fr/verification/

## Twitter data

We collected two different datasets using the Twitter API: a collection of tweets citing French media news stories and the network of Twitter users who follow French parliamentarians.

The selection process of our corpus of tweets is mirroring the composition of the top layer by using the open source tool Gazouilloire (Ooghe-Tabanou et al. 2021) that leverages both Twitter's live streaming and search APIs. Since May 2018, we have been running a continuous collection of every single tweet that cites URLs published by each of the media outlets present in our initial perimeter of media websites. Our dataset D2 has amounted to nearly 100M tweets

We also collected a large network composed of the list of followers of French representatives in the two chambers of the French Parliament, the Senate and the Assemblée nationale. Our data collection starts with the set of 831 French parliamentarians (later referred as MPs) present on Twitter, assembled, and curated continuously by NGO Regards Citoyens[5]. We then proceeded to collect all the followers of the accounts of MPs. This collection was conducted in May 2019, and resulted in the constitution of a corpus of followers that amounts to nearly 4.5 million unique Twitter accounts. We then removed followers following less than 3 MPs, and again removed users which had a repeated set of followed MPs. We finally obtained a bipartite network connecting 368.831 Twitter accounts to 831 MPs.

The construction of such a social network allows us in turn to follow the methodology described by Barbera et al. (2015) for inferring ideological features through the multidimensional scaling of the graph of the MPs and their followers (that we will describe later). The dataset D3 therefore consists in the inferred ideological positioning of nearly 370k French Twitter users in a 2d latent space. Twitter consequently provides an indirect way to estimate the degree of polarization of French Twitter as users' positions in the ideological space can be used in turn to locate individual media outlets and analyze the ideological distribution of French media.

## Facebook data

Finally we are also interested in the consumption of news stories on Facebook. To estimate how popular they are in terms of clicks, views and shares on the social network, we used the Condor dataset[6]. The Condor dataset provides aggregated statistics about every URL shared on Facebook that attracted more than 100 likes (over a period of 31 months) ranging from January 2017 to July 2019. A system of differential privacy was added to the original statistics. However, the noise level is documented for each variable and low enough such that the estimates for the popularity of sources are reliable. This is dataset D4.

---

[4] https://github.com/medialab/corpora/tree/master/polarisation/papers/2021-IC2S2
[5] https://github.com/regardscitoyens/twitter-parlementaires
[6] We obtained access to the dataset through the Social Media and Democracy Research Grant "I read it on Facebook" https://www.ssrc.org/fellowships/view/social-media-and-democracy-research-grants/grantees/cointet/

# Results

## Hyperlinks network analysis

The hyperlinks network (D1) classically exhibits a high level of heterogeneity in terms of credit distribution. Certain sources from the traditional print media still receive a large amount of authority (in the form of incoming citations) from many sources, while certain sources are almost never cited. Among the most central nodes when ranked by PageRank are prominent national broadsheet newspapers (Libération, Le Monde, Le Figaro), reference magazines (Télérama, Le Point, L'Express), prominent regional press (Ouest France, La Dépêche du Midi) or main TV/radio stations (RFI, RTL, BFM/TV).

In order to characterize the more refined structural properties of the hyperlinks network, we use a bayesian variant (Funke and Becker, 2019) of the Stochastic Block Model as introduced by Peixoto (2014), which aims at extracting the meso-level structure of a network. A Stochastic Block Model (later referred as an SBM) is a generative model that provides a statistically minimal description of the topology of a network while avoiding both overfitting and underfitting. The end result of a SBM is not only a (hierarchical) partition, but also a set of probabilities that describe the likelihood for linking a node of one group to another group. Peixoto's SBM allows identifying sets of nodes that share the same citing and cited profile towards nodes grouped in other blocks. Note that the blocks which are inferred by SBM may gather assortative sets of nodes, but also more complex disassortative structures such as sets of nodes citing nodes from another block and receiving links from a third group. Additionally, SBM builds a hierarchical representation of the blocks that provide a high-level description of the connectivity patterns of each individual media outlet. We use the degree corrected version of the model that allows us to build blocks, which potentially gather nodes ranging over large degree intervals. Figure 2 proposes a visual representation of the network and its hierarchical block structure, that we annotated with our own labels.

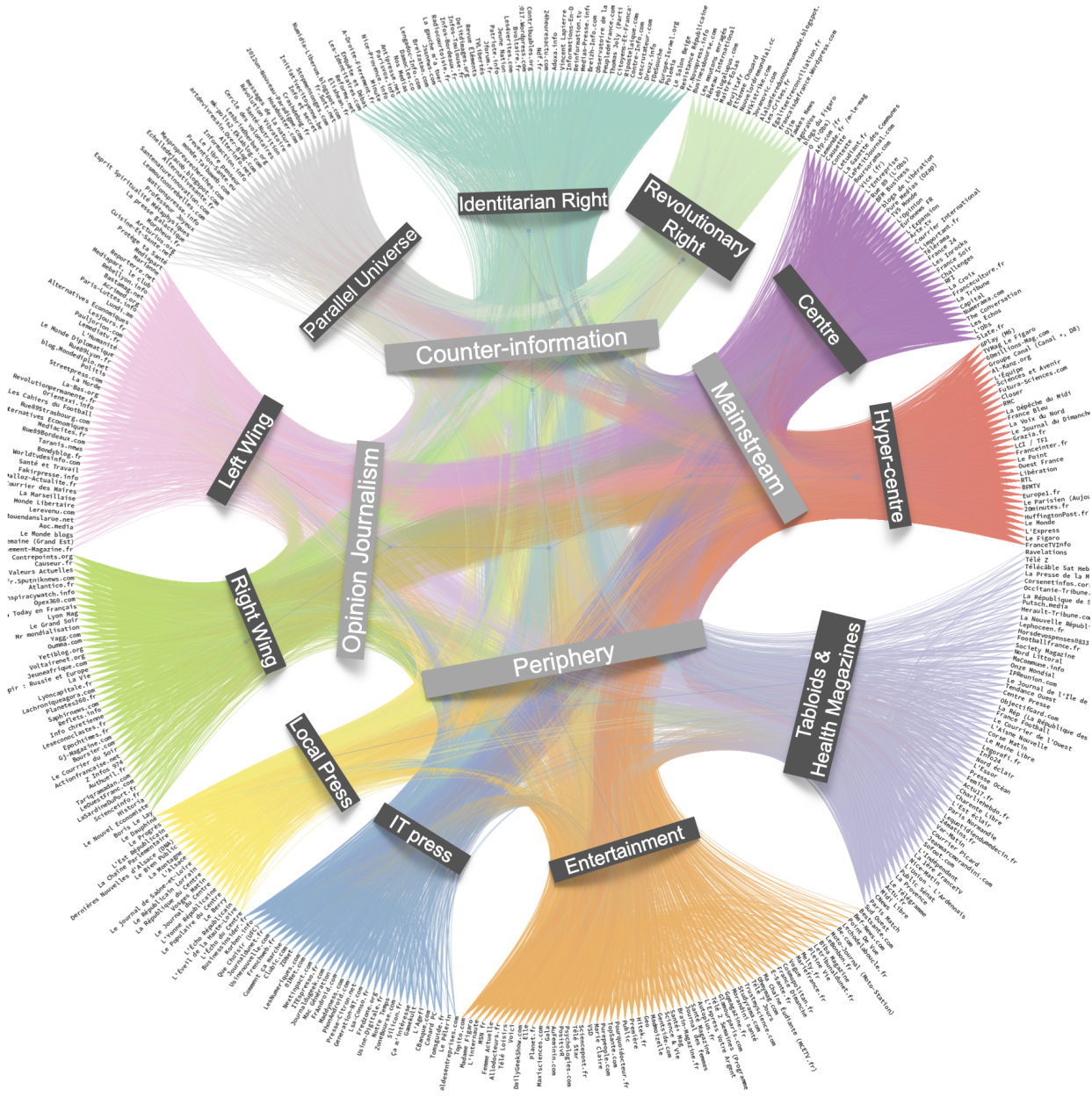

**Fig 2.** The hyperlinks network connecting media outlets is structured in a two level block structure. At the coarser grain, the 4 emerging groups of blocks have been interpreted as: Mainstream media, Counter-informational space, Opinion Journalism and the Periphery.

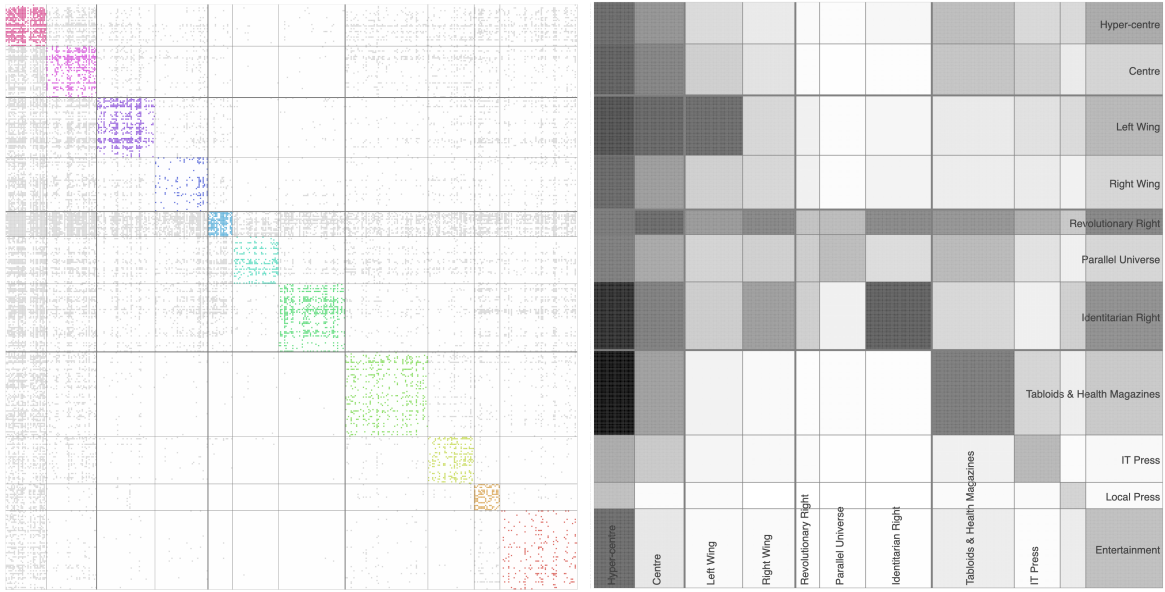

**Fig 3.** [Left] The empirical adjacency matrix shows the presence/absence of links between two media in our corpus. [Right] Visual representation of the associated generative SBM model. Blocks were labelled manually. Darker (i,j) rectangles code for a higher likelihood of hyperlinks connecting block i to block j. From this representation it is quite obvious that media from the hyper-centre receive links from everywhere else when counter-informational blocks cite every other block while hardly getting attention from any other block.

A possible representation of the hierarchical block structure is shown above. We plotted the binary adjacency matrix showing every potential link between two media outlets (Fig. 3, left). Sources have been pre-ordered so that media outlets belonging to the same block are adjacent in the matrix. 11 blocks (identified by the color of their diagonal cells) have been automatically extracted. The inferred hierarchical model shows that these 11 elementary blocks can be clustered in 4 higher-level blocks that we delimitate visually using thicker lines. The table 1 below provides basic statistics about each block along with the higher-level block they belong to: *Mainstream media*, *Opinion journalism*, *Counter-informational space* or *Periphery*. The simplified adjacency matrix on the right summarizes the connectivity patterns captured by the model at the lowest block model. We also included hand-p labels for each of the 11 blocks. Darker rectangles at the intersection between two blocks i and j mean higher probability for a media in block j to be cited by a media in block i.

| Continents | Blocks | Number of media | Median in-degree | Median PageRank | Median out-degree | Eigenvector centrality | Median Year of creation |
|---|---|---|---|---|---|---|---|
| Mainstream media | Hyper-centre | 30 | 207 | 10.79 | 58 | 0.66 | 1971 |
|  | Centre | 37 | 93 | 3.27 | 41 | 0.31 | 1992 |
| Opinion Journalism | Left Wing | 43 | 27 | 1.02 | 39 | 0.08 | 2007 |
|  | Right Wing | 39 | 29 | 0.67 | 37 | 0.04 | 2008 |
| Counter-informational space | Revolutionary Right | 18 | 35 | 0.66 | 215 | 0.04 | 2008 |
|  | Identitarian Right | 49 | 21 | 0.51 | 52 | 0.03 | 2010 |
|  | Parallel Universe | 34 | 15 | 0.50 | 51 | 0.02 | 2012 |
| Periphery/Specialized media | Local Press | 19 | 39 | 2.27 | 24 | 0.13 | 1944 |
|  | IT Press | 34 | 30 | 1.33 | 21 | 0.09 | 2000 |
|  | Tabloids & Health Magazines | 61 | 32 | 1.14 | 26 | 0.08 | 2000 |
|  | Entertainment | 57 | 26 | 1.06 | 12 | 0.06 | 1980 |

**Tab 1.** Aggregate statistics of media blocks: we compute several topological features aggregated over each block: median in-degree, out-degree, PageRank, eigenvector centrality. We also manually identified the date of creation of each source to measure the median year of creation of each block.

Our first observation is that French media producers form a highly structured pyramid, with most of the hyperlinks pointing toward the two blocks (the *Centre* and the *Hyper-center*) that form what we call the *"mainstream media"* at the higher level. These two blocks receive attention (translated as inbound links) from the entire media ecosystem. Their eigenvector centrality and their PageRank is very high in comparison to the other blocks. They reciprocate in a contrasted manner, hardly giving any credit (at least in the form of hyperlinks) to the *Counter-informational space*, but citing quite generously the *Periphery*. The two groups of the *Centre* and the *Hyper-centre* gather the main French media of the press, radio and television (Le Monde, Le Figaro, Europe 1, BFM TV). Almost all of these publications were introduced before the digital revolution, have the largest audience in the French media space, and are able to capitalize on the authority they have over the French public. Even if their political orientations may vary from centre-left to centre-right, they appear together in the classification as the media most cited by the others, regardless of their political positions. The practices and inner conventions followed by any journalist in the newsrooms explain how the French media system achieves such a high level of gate-keeping. *Mainstream media* observe a strict discipline when discussing information published in the *counter-informational space*. For instance, when evaluating questionable information, fact-checkers are likely to use screenshots to refer to a questionable source rather than share its domain name or make a Twitter account readily clickable.

*Opinion journalism* is the second higher-level collection of blocks: it gathers websites which defend a strong political line. On the fringes of the *mainstream media*, *opinion journalism* is divided into two clearly separated sub-spaces: the right-wing (Causeur, Valeurs actuelles, Atlantico, etc.) and the *left-wing* *media* (Mediapart, L'Humanité, Le monde diplomatique, etc.). The politicized younger media, in many cases exclusively digital, is exerting critical pressure on the *mainstream media* continent. These *right* and *left* publications are sufficiently integrated into the *mainstream media* space to receive, both right and left, a fair number of links from the *Hyper-centre* and *Centre* blocks of the *mainstream* space. They are not marginalized and constitute the political fringes of the French public sphere. However, there is a striking structural asymmetry between the *left-wing* and the *right wing* sub-spaces if we pay attention to the matrix of links exchanged between the different blocks (Fig. 2). *Left wing media* seem more self-centered citation-wise, and they also receive much less attention than *right wing media* from the *counter-informational space*. This difference between the two groups highlights an important structural fact: *right-wing media* could be a gateway between the *mainstream media* and the *counter-informational space* that more frequently produces dubious or manipulative information. Nevertheless, the French situation is not at all comparable to the divided US mediaspace. The right-wing media receive many links from the identitarian far-right outlets, but address few links in their direction.

*Counter-informational space* has the most striking connectivity pattern as it hardly receives any links from the rest of the ecosystem even though it sends a lot of citations to the entire ecosystem. It clearly illustrates how the most legitimate part of the media ecosystem purposely ostracizes media outlets sharing misinformation or tries to impose its propaganda. A closer look at the counter-informational space reveals three blocks which we labeled the *Parallel Universe*, *Revolutionary right*, and *Identitarian Right* blocks. The first sub-space gathers websites that suggest the existence of alternative worlds. Two families of publications emerge from this heterogeneous set: sites devoted to alternative health, and those defending spurious theories based on supernatural effects or large-scale conspiracies. These sites are recent, have a confidential audience and a very low PageRank, even if the themes they promote, especially on health issues, have a much higher popularity on Facebook. The other two sub-spaces match a traditional far-right split in France. The group of the *identitarian* far-right combines the most traditionalist components of the French extreme right with the most radical ethno-racial identity fringes. It is dominated by one of the most influential historical sites of the French extreme right, FDeSouche (which stands for "Français de souche") founded in 2006. Mixing ancient traditions of the extreme right (catholic, conservative, royalist), these websites are characterized by their hostility to sub-Saharan and Muslim immigrant populations. The *Revolutionary Right* sub-space distinguishes itself from the *Identitarian Right* by the centrality given to conspiracy and the domination of a World Order. The themes it promotes are borrowed from both the right and the left to denounce the power of the elites, American imperialism, and to accuse Freemasonry, the Illuminatis or the Jewish conspiracy. Egalité et réconciliation, funded by Alain Soral, dominates a thriving galaxy of small websites.

Finally the *Periphery* attracts less attention overall and proportionally less from *opinion journalism*. This cluster of media brings together different types of specialized media: Local Press, Technology press, Entertainment Magazines and Health Magazines & Tabloids. When citing media outlets outside, it is mostly toward the *Hyper-centre*. Some of these media have a large audience but are significantly less important in news coverage and political information.

The placement of professional media from both the right and the left at the top of the hierarchy of authority of hypertext links demonstrates the permanence of the group of central media in the French information landscape. In spite of the promise of an upheaval of the media industry with the arrival of new actors, outlets that dominate the media space existed before the development of the web. The web has not reshuffled the cards in the information sector to transform the hierarchy of legitimacy. Even if the web has indeed created specificities that will appear in the other subspaces of our cartography, a process of standardization has allowed, in the French case at least, a rather traditional renewal of the media hierarchy (Napoli, 2019 ; Hindman, 2018).

## News diet using social media data

The previous section shows a clearly cut picture of the structure of the media production ecosystem with counter-information space being ostracized by more legitimate outlets. We now turn to social media news stories consumption statistics and try to quantify how much attention French users are giving to each subspace online. We leverage our dataset, compiling every single tweet over a one year period (D2) that cites a URL pinpointing to a news story from one of our media . We have plotted these statistics on the histogram Figure 4.

We also plot certain aggregated topological features from the hyperlinks network. We observe that the out-degree of blocks is quite homogeneous across the blocks. As we already observed in the previous section, the highly hierarchical structure of the media hyperlinks network only shows in the uneven distribution of in-coming links, which flow largely towards *Mainstream media* (more than 50%). Conversely, *counter-informational space* attracts less than 10% of the incoming links, which is much less than the *Periphery*, or *Opinion Journalism*. One should also keep in mind that most counter-informational links are actually endogenous.

When looking at statistics drawn from Twitter, we observe that the unevenness of visibility is even stronger. Nearly two thirds of the tweets citing media outlets are citing *mainstream media*. Less than 5% of tweets are citing counter-informational media. Interestingly, the share of Twitter users who account for those tweets is even smaller (around 2%). These results show that French Twitter users do not circulate a larger share of links from peripheral or alternative media or from sources that may produce questionable information. In fact, the opposite phenomenon is observed. When we take a "distant" look at the French media ecosystem, Twitter influencers seem to focus even more on the most popular and central newspapers of the professional journalism space.

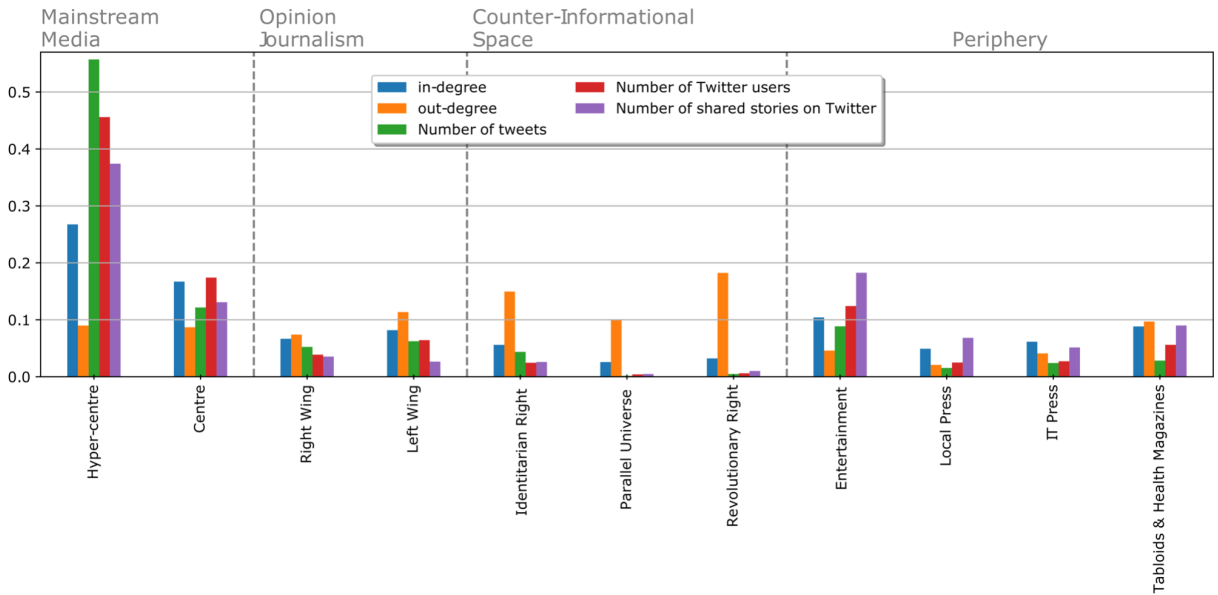

**Fig 4.** Distribution of attention of the 11 different blocks, divided by continent, measured in terms of incoming/outgoing links on websites, and activity on Twitter. We distinguish between the number of tweets, the total number of distinct Twitter users and the total number of distinct news stories.

One may think this pattern is specific to Twitter, as the population of the social network is quite biased in favor of young, educated and wealthier segments of the population (Barbera, Rivero, 2015; Boyadjian, 2014). We leverage the D4 dataset to measure not only the number of times a French media URL has been shared on Facebook, but also the number of times it has been liked, commented or clicked on when it is displayed in users' news feed. The Condor dataset (see Fig 5) shows that the news consumption pattern on Facebook does not depart from the one observed on Twitter. *Mainstream media* still attracts more than half of the total activity. *Identitarian Right* content and *Counter-informational* space at large generate slightly more activity on Facebook than on Twitter. However, despite the largest share of shares, the number of clicks that really capture the number of people who get exposed to news from those websites is really limited (3%).

This should not come as a surprise: the largest discrepancy we observe between the level of activity on Twitter and Facebook concerns our *Tabloïds and Health Magazine* category, which generate a very large portion of clicks on Facebook when compared with Twitter. These results have been observed in other European countries (Flechter & al., 2018). When observed on a large scale, on Twitter like on Facebook, the structure of information shared by users reproduces the traditional media hierarchy. However, this overall statistical result may hide more local phenomena of news dissemination that stems from the edges of the media space defined in our corpus.

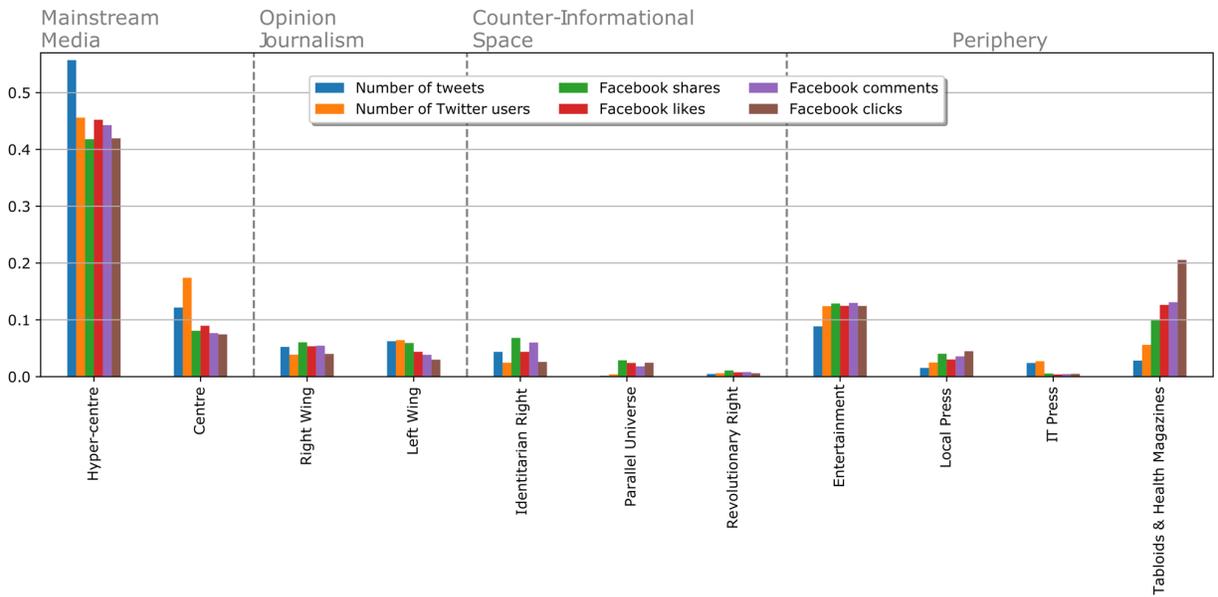

**Fig 5.** Distribution of the attention of the 11 different blocks, divided by continent, on Twitter and Facebook. We distinguish between the number of shares, likes, comments and clicks.

# Looking for ideological polarization

The topological analysis of the hyperlinks network connecting media sources, and the analysis of the popularity of sources online, suggests that we do not observe the same level of polarization in the French media ecosystem as the one observed in the US : Benkler et al. (2019) shows how asymmetric the ideological distribution of American media is with a clear concentration of high visibility news outlets on the extreme right side of the political spectrum. Our inspection of social media shows that mainstream media (which is politically diversified but never extreme) attracts most of the audience. However the structural decomposition offered by the stochastic block model does not clearly map a divide between the right and the left wings. To try to map our analysis of such a traditional political divide, we follow the methodology introduced by Barbera (2015). We infer the ideological score of a series of nearly 368 thousand individuals who follow at least 3 representatives out of the 831 French MPs who have an active Twitter account. The resulting embedding is illustrated in Fig 6 below. The precise method and the description of the embedding and its inferred axes are detailed in Cointet et al. (2020). Put simply, the original network connecting MPs to their followers is analyzed as a binary adjacency matrix. We produce a reduced-dimensionality representation using a Correspondence Analysis (CA). Using the Chapel Hill Expert survey (Bakker et al., 2020), we show that the second principal component aligns with a left-right economics axis (Ramaciotti Morales et al., 2020). The rest of our analysis exclusively relies on this dimension that we will simply refer to as "ideological score" for the sake of clarity. Correspondence analysis distributes the ideological score of Twitter accounts in an uneven way. This is the reason why we later binned the right and left positions of the entire set of MPs' followers (368 thousand Twitter users) in nine evenly populated bins ranging from extreme left to extreme right. The same bins are later used to plot the ideological distribution of news stories Fig. 7.

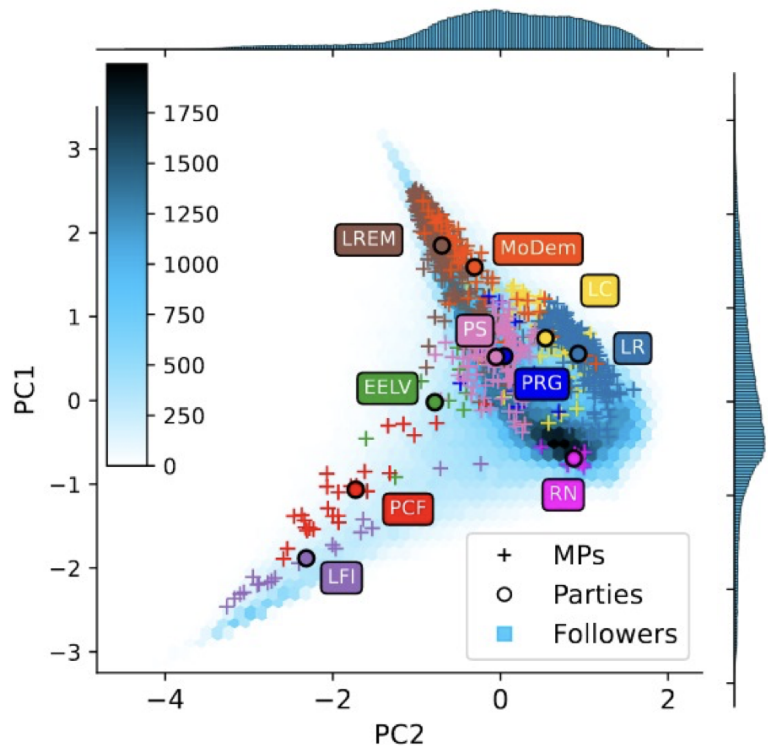

**Fig 6.** Ideological scaling of the French MPs followers/followees network. The second principal component clearly distributed parties and Twitter user accounts on the right/left political dimension. Representatives are colored according to the party they belong to. The colored circles correspond to the average position of MPs in each of the main 10 French party.

This embedding allows, in turn, to measure the average ideological score of URLs shared on Twitter (which can be approximated since the average ideology of its sharers), and how visible news stories are according to their inferred ideological orientation. Here we simply computed, using the dataset D2, the ideological distribution of the overall set of news stories shared on Twitter. More precisely, for each news story, we were able to extract the ideological distribution of the Twitter users who shared this URL and were part of D3. Adding distributions for every news story published by any of our media sources over a one year period ,we can plot the blue distribution Fig 7. We observe that a larger number of tweets originate from either extreme left or extreme right users, which is consistent with prior results obtained in Spain (Barbera & Rivero, 2015). Interestingly, we plotted the same distribution in orange where the contribution of each individual user was weighted by its number of followers. This offers a rough estimate of the reach of their tweet. The normalization operations give us a better image of the kind of political landscape one can observe on Twitter. This distribution is clearly not bi-modal, challenging the idea that the digital French public space could be split in two irreconcilable subspaces. Here, most of the feed a "new" user on Twitter is likely to be exposed to is coming from users with the least extreme political preferences.

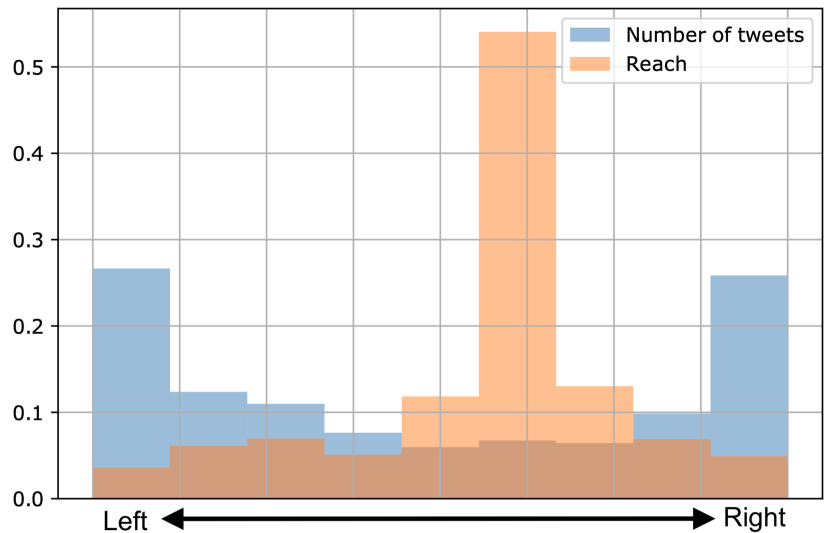

**Fig 7.** Distribution of the political orientation of Twitter users sharing news. Our 368k users are first uniformly distributed over 9 bins. We then measure the aggregated distribution of all the news stories that are shared on Twitter. The blue bimodal distribution indicates that more extreme users (on both sides) are likely to produce much more tweets than more moderate users. However, when weighting the distribution with the number of followers of each Twitter user, the global picture looks much different, with a more traditional Poisson distribution centered at the center of the political spectrum.

# Discussion and Conclusion

In France, there is a significant number of fake news producers or producers of news that conveys extreme political messages (Cordonier, Brest, 2021). On many occasions, fact-checkers of professional media (Decodex at Le Monde, CheckNews at Libération, Factuel at AFP...) have highlighted the circulation of false information. However, in the French case, this information rarely succeeds in occupying the center of the media agenda. This article proposes a structural explanation to this phenomenon. The French media space has maintained a hierarchical and pyramidal structure despite the digitization of most information producers. The editorial staff of the main French media have very different editorial points of view and hold contrasting political positions. However, the space of professional journalists shares the same conceptions of the ethics of information and exercise a critical eye on other media that makes it difficult to put into circulation obviously erroneous information. This configuration differs significantly from the American situation. This result supports a finding of studies on disinformation: the polarization of the journalistic space and the circulation of fake news are phenomena that only become more widespread when dominant and influential actors in the political/journalistic space relay themes and dubious content from the fringe of the information space (Heiberger & all., 2021; Evanega & al., 2020; Nielsen, 2020).

The ecological perspective implemented in this article shows the interest of a comparative approach. Media spaces in different countries have different histories and modes of organization. Contemporary debates on filter bubbles, polarization or disinformation are characterized by a certain form of technological determinism that makes social networks the sole cause of new information disorders. A comparative approach shows that it is necessary to articulate a sociological approach of the media to the understanding of the effects of the new digital infrastructures.

# Acknowledgement

This work has been funded by the French National Agency for Research under grant ANR-19-CE38-0006: Geometry of Public Issues (GOPI). It also builds on data from the "I read it on Facebook'' project of the Social Science Research Council's Social Science One programme. It also took advantage of the Hyphe and Gazouilloire softwares developed at médialab thanks to the support of DIME-Web, part of DIME-SHS research equipment financed by the EQUIPEX program (ANR-10-EQPX-19-01). We are also grateful to Paul Girard, Tim Fingerhut and Audrey Baneyx for their help in curating our list of media websites. In compliance with the General Data Protection Regulations 2016/679, the research project Ideology scaling Twitter France, of which the data were used, was declared on March 19th 2020 to Sciences Po's data processing register.

# Appendix

The list of 421 media as categorized in the various blocks:

| MAINSTREAM MEDIA | OPINION JOURNALISM |
| --- | --- |

| Hyper-centre | Centre | Right Wing | Left Wing |
|---|---|---|---|
| LCI / TF1 | Courrier International | Russia Today en Français | Mediacites.fr |
| Le Monde | L'Opinion | Valeurs Actuelles | Bondyblog.fr |
| RMC | La Tribune | fr.Sputniknews.com | Revolutionpermanente.fr |
| FranceTVInfo | Arte.tv | Opex360.com | Reporterre.net |
| HuffingtonPost.fr | France 24 | Lyon Mag | Mediapart |
| 20minutes.fr | Les Inrocks | Jacques Sapir | Lundi.am |
| Closer | Télérama | Info chrétienne | Acrimed.org |
| France Bleu | TV5 Monde | Oumma.com | Mediapart, le club |
| Le Parisien | Challenges | Leseconoclastes.fr | La-Bas.org |
| RTL | RFI | Causeur.fr | Taranis.news |
| Le Figaro | Les Echos | Voltairenet.org | Le Monde Diplomatique |
| La Voix du Nord | Franceculture.fr | Lachroniqueagora.com | Alternatives Économiques |
| Europe1.fr | France Soir | Atlantico.fr | L'Humanité |
| BFMTV | Slate.fr | Contrepoints.org | Politis |
| Ouest France | blogs de Libération | Conspiracywatch.info | Fakirpresse.info |
| Le Journal du Dimanche | Pure Medias (Ozap) | Yagg.com | Streetpress.com |
| L'Équipe | La Croix | Planetes360.fr | Rue89Lyon.fr |
| Al-Kanz.org | Capital | Le Grand Soir | Rebellyon.info |
| Grazia.fr | Causette | Reflets.info | Bastamag.net |
| L'Express | L'Entreprise | Saphirnews.com | Rue89Strasbourg.com |
| Le Point | Afp.com /fr | Scienceinfo.fr | Paris-Luttes.info |
| Sciences et Avenir | L'Obs | Tariqramadan.com | blog.Mondediplo.net |
| Franceinter.fr | The Conversation | Epochtimes.fr | blogs Alternatives Éco |
| Libération | Rue 89 (L'Obs) | Gj-Magazine.com | Santé et Travail |
| La Dépêche du Midi | Limportant.fr | Yetiblog.org | Monde Libertaire |
| Futura-Sciences.com | Lemonde.fr /m-le-mag | Le Nouvel Economiste | Lemediatv.fr |
| Groupe Canal + | L'Expansion | Boursier.com | Marianne |
| TVMag Le Figaro | BFM Business | Mr mondialisation | Orientxxi.info |
| 60millions-Mag.com | Letudiant.fr | Le Courrier du Soir | Le Courrier des Maires |
| 6Play (M6) | Boursorama.com | La Vie | La Horde |
| | Numerama.com | Jeuneafrique.com | Rouendanslarue.net |
| | Vice (fr) | Actionfrancaise.net | Rue89Bordeaux.com |
| | La Gazette des Communes | Boris Le Lay | Worldtvdesinfo.com |
| | O (L'Obs) | Authueil.fr | Lesjours.fr |
| | Contexte | LeOuestFranc.com | La Marseillaise |
| | LePetitJournal.com | LaSardineDuPort.fr | Les Cahiers du Football |
| | Euronews FR | Lyoncapitale.fr | Dalloz-Actualite.fr |
| | | Z Infos 974 | Aoc.media |
| | | Historia | La Semaine (Grand Est) |
| | | | Lerevenu.com |
| | | | Pauljorion.com |
| | | | Environnement-Magazine.fr |

|  |  |  | Le Monde blogs |
| --- | --- | --- | --- |

| | COUNTER-INFORMATIONAL SPACE | |
| --- | --- | --- |
| **IDENTITARIAN EXTREME RIGHT** | **REVOLUTIONARY RIGHT** | **ALTERNATIVE HEALTH AND UFOs** |
| Contre-Info.com | Etienne Chouard | Stopmensonges.com |
| Lengadoc-Info.com | AgoraVox | Lesbrindherbes.org |
| TVLibertés | francaisdefrance.Wordpress.com | Esprit Spiritualité Métaphysiques |
| Adoxa.info | Egaliteetreconciliation.fr | Alternativesante.fr |
| La gauche m'a tuer | blogs du Figaro | Informaction.info |
| Thomas Joly (Parti de la France) | Les-Crises.fr | mk-polis2.Eklablog.com |
| Dreuz.info | Ojim | Numidia-Liberum.blogspot.com |
| Ripostelaique.com | Jovanovic.com | messages de la nature |
| Ndf.fr | Fawkes News | Elishean.fr |
| 24heuresactu.com | Wikistrike.com | Crashdebug.fr |
| Parisvox.info | Alalumieredunouveaumonde | Révolution Vibratoire |
| Radiocourtoisie.fr | Nouvelordremondial.cc | Info et secret |
| Les4verites.com | Leblogalupus.com | Initiativecitoyenne.be |
| Infos-Bordeaux.fr | Maitre-Eolas | Echelledejacob.blogspot.com |
| ns2017.Wordpress.com | Brujitafr | 2012un-Nouveau-Paradigme.com |
| Reinformation.tv | Les moutons enragés | Lesmoutonsrebelles.com |
| Delitdimages.org | Businessbourse.com | Santé-Nutrition |
| Le Salon Beige | Réseau International | Prevention-Sante.eu |
| Résistance Républicaine | | Morpheus.fr |
| Observatoire de la Christianophobie | | La presse galactique |
| Revue Éléments | | artdevivresain.Over-Blog.com |
| Informations-En-Direct-France. | | Arcturius.org |
| Contribuables.org | | fr.Sott.net |
| A-Droite-Fierement.fr | | Santenatureinnovation.com |
| Infos-Toulouse.fr | | monde.Taibaweb.com |
| Citoyens-Et-Francais.fr | | Professeur Joyeux |
| Breizh-Info.com | | Cercle des volontaires |
| Europe-Israel.org | | Alterinfo.net |
| Damocles.co | | Le libre penseur |
| Polemia | | Hoaxbuster.com |
| FDeSouche | | Protège ta Santé |
| Peupledefrance.com | | Nationspresse.info |
| Minute | | Cuisine-Et-Sante.net |
| Medias-Presse.info | | Mespropresrecherches.com |
| Patriote.info | | |
| Vincent Lapierre | | |
| Nice-Provence.info | | |
| Bvoltaire.fr | | |
| fr.Novopress.info | | |
| Les-Identitaires.com | | |
| Lescrutateur.com | | |
| Enquête et Débat | | |
| Jssnews.com | | |
| Nos Médias | | |

| | | | |
|---|---|---|---|
| Breizatao.com | | | |
| Antipresse.net | | | |
| Jeune Nation | | | |
| Reforme.net | | | |
| Jforum.fr | | | |

| | SPECIALIZED MEDIA | | |
|---|---|---|---|
| **Regional press** | **Technology** | **Local Press** | **Health, beauty & tabloïds** |
| L'Yonne Républicaine | Gamekult | Corse Matin | Brain-Magazine.fr |
| L'Alsace | Macgeneration | La 1ère FranceTV | Télé Star |
| Dernières Nouvelles d'Alsace (DNA) | Maddyness.com | Presse Océan | Marie Claire |
| Le Populaire du Centre | Generation-NT.com | 24matins.fr | Télé Loisirs |
| Vosges Matin | 01Net.com | Actu.fr | Journal des femmes |
| Le Berry | ZDNet | Midi Libre | Voici |
| La Montagne | Tomsguide.fr | Le Courrier de l'Ouest | Télé 7 Jours |
| La République du Centre | Usinenouvelle.com | L'Indépendant | Gala |
| Le Bien Public | FrAndroid.com | Nice-Matin | Mariefrance.fr |
| L'Écho Républicain | Businessinsider.fr | Var-Matin | Femme Actuelle |
| Le journal de Saône-et-Loire | Fredzone.org | Actu17.fr | Purepeople.com |
| Le Républicain Lorrain | Journaldunet.fr | L'Union - L'Ardennais | Public |
| L'Est Républicain | PhonAndroid.com | Centre Presse | Cosmopolitan.fr |
| Le Progrès | ITEspresso.fr | Le Maine Libre | Vogue |
| Le Journal du Centre | LesNumeriques.com | Nord éclair | Madame Figaro |
| L'Éveil de la Haute-Loire | Presse-Citron.net | L'Est éclair | VSD |
| Le Dauphiné | Clubic.com | Jeanmarcmorandini | TopSante.com |
| L'Écho du Centre | Journaldugeek.com | Le Télégramme | Glamourparis.com |
| La Chaîne Parlementaire | Frenchweb.fr | Courrier Picard | AuFeminin.com |
| | Usine-Digitale.fr | Info24 | Santé+ Mag |
| | Korben.info | ObjectifGard.com | Biba Magazine |
| | CBanque.com | L'Aisne Nouvelle | Gqmagazine.fr |
| | ZoneBourse.com | La Nouvelle République des Pyrénées | Télé 2 Semaines |
| | Lsa-Conso.fr | La Provence | Allodocteurs.fr |
| | Comment √ßa marche | Charente Libre | Elle |
| | Silicon.fr | Sofoot.com | Première |
| | Lejournaldesentreprises.com | Lephoceen.fr | France Dimanche |
| | Notre Temps | Onze Mondial | Be.com |
| | √áa m'intéresse | L'Essor | Psychologies.com |
| | Nextinpact.com | IPReunion.com | Pleine Vie |
| | L'Agefi | La République de Seine-et-Marne | Santé Magazine |
| | Le Pèlerin | La Presse de la Manche | Letribunaldunet.fr |
| | Canard PC | Tendance Ouest | E-Sante.fr |
| | Que Choisir (UFC) | Charliehebdo.fr | Ma Chaîne etudiante |
| | | Herault-Tribune.com | Autoplus.fr |
| | | La Rép (La République des Pyrénées) | Pourquoidocteur.fr |
| | | Paris Normandie | Ohmymag.com |
| | | Le Journal de l'√éle de la Réunion (Clicanoo.re) | LeBonbon.fr |
| | | Lequotidiendumedecin.fr | Melty.fr |
| | | Nord Littoral | Morandini santé |
| | | Society Magazine | Studyrama.com |
| | | MaCommune.info | Point De Vue |
| | | Occitanie-Tribune.com | Geo |
| | | Footballfrance.fr | L'Express Votre Argent |
| | | Corsenetinfos.corsica | Madmoizelle |
| | | Télé Z | Bestsante.com |
| | | France Football | Moto-Journal |
| | | Putsch.media | Bmf-News.com |
| | | Ravelations | Lechodelaboucle.fr |
| | | Public Sénat | Gentside.com |
| | | Legorafi.fr | Planet.fr |
| | | Télécâble Sat Hebdo | L'internaute |
| | | Horsdevospenses | Science & Vie |
| | | Sud Ouest | Trustmyscience.com |
| | | Femina | DailyGeekShow.com |
| | | CNews | Hitek.fr |
| | | Paris Match | Maxisciences.com |
| | | | PositivR |
| | | | MSN fr |
| | | | Topito.com |
| | | | Sciencepost.fr |